# Characterization of precipitation in Al-Li alloy AA2195 by means of atom probe tomography and transmission electron microscopy

Muna Khushaim*, Torben Boll*, Judith Seibert**, Ferdinand Haider**, Talaat Al-Kassab*

*Physical Sciences and Engineering Division, King Abdullah University of Science and Technology (KAUST), Thuwal 23955-6900, Kingdom of Saudi Arabia.

** University Augsburg, Inst. f. Physics, D – 86159 Augsburg, Germany.

Corresponding author: Muna Khusahim, Tel:±966503372178, Fax: +966128025322, email: Muna.Khushaim@kaust.edu.sa

**Abstract:**

The microstructure of the commercial alloy AA2195 was investigated on the nano-scale after conducting a T8 tempering. This particular thermo-mechanical treatment of the specimen resulted in the formation of platelet-shaped $T_1(Al_2CuLi)/\theta'(Al_2Cu)$ within the Al-matrix. The electrochemically prepared samples were analyzed by scanning transmission electron microscopy and atom probe tomography for chemical mapping. The $\theta'$ platelets, which are less than 2 nm thick, have the stoichiometric composition consistent with the expected $(Al_2Cu)$ equilibrium composition. Additionally, the Li distribution inside the $\theta'$ platelets was found to equal the same value as in the matrix. The equally thin $T_1$ platelet deviates from the formula $(Al_2CuLi)$ in its stoichiometry and shows Mg enrichment inside the platelet without any indication of a higher segregation level at the precipitate/matrix interface. The deviation from the $(Al_2CuLi)$ stoichiometry cannot be simply interpreted as a consequence of artifacts when measuring the Cu and Li concentrations inside the $T_1$ platelet. The results show rather a strong hint for a true lower Li and Cu contents and hence, supporting reasonably the hypothesis that the real chemical composition for the thin $T_1$ platelet in the T8 tempering condition differs from the equilibrium composition of the thermodynamic stable bulk phase.

1. Introduction:

One of the upcoming materials used for the structural components of aircrafts are aluminum-lithium alloys. This class of alloys is commonly selected because it exhibits an enhanced strength/weight ratio that makes it an attractive material for weight critical applications. A series of Al-Li-Cu alloys with minor amount of Mg, Ag and Zr known as Weldalite that exhibit desirable combinations of strength and toughness have been developed in recent years [1].

The current study focuses on one member of the Weldalite family, namely the commercial AA2195 alloy, which is considered as a promising candidate for aerospace applications due to its high strength, high modulus, low density, good corrosion resistance, excellent fatigue properties and fracture toughness at cryogenic temperatures [2]. The alloy exhibits superior mechanical properties after processing a thermo-mechanical treatment termed 'T8'. This treatment begins with a solution treatment in a single phase region, followed by quenching to obtain a supersaturated solid solution. Plastic deformation (cold working) is then applied to the samples. The final stage is an artificial aging far below the temperature of the strengthening phase solvus line [3]. The presence of various micro-alloying elements at a variable volume fraction and adequate subsequent treatment causes the precipitation of different possible phases within the Weldalite family such as $T_1(Al_2CuLi), \theta'(Al_2Cu), \beta'(Al_3Zr)$ and $\delta'(Al_3Li)$. The $T_1$ phase has a plate-shaped morphology lying on {111} planes and is the primary phase responsible for the strengthening effect in AA2195 at temperatures below 533 K (260 °C) [4]. The nucleation of this phase is difficult without the presence of secondary alloying elements such as Mg and Ag, which can offer nucleation sites within the matrix for this phase [5]. The nucleation of the $T_1$ phase also depends on the application of plastic deformation prior to an artificial aging [6]. The $T_1$ phase co-exists with the $\theta'(Al_2Cu)$ phase which lies as plates on the

{100} matrix planes [7]. This phase is dominant in the early stage of the heat treatment at elevated temperatures 588 K (315 °C) to 644 K (371 °C) [4]. In addition to the $T_1$ and $\theta'$ phases, Zr forms coherent spherical $\beta'(Al_3Zr)$ precipitates with $L1_2$ - structure that can already form during the alloying process within the melt. They are designed to control the grain structure and degree of recrystallization after the solidification of the ingot [8]. Finally, the $\delta'(Al_3Li)$ is usually reported to precipitate upon quenching from the melt for alloy compositions higher than 5 at.% Li in the matrix [9,10,11]

Atom probe studies have provided a range of information on the different types of precipitates present in Al-Li-Cu alloys. Murayama and Hono [12] reported on their observation of Mg clusters at the early stage of aging and on the possibility of the segregation of Mg and Ag atoms to the broad interface of the $T_1$ plates. Moreover, Gault et al. [13] studied the early stages of clustering, precipitate interactions, solute segregation at the matrix/precipitate interfaces, and the chemical composition of the different phases.

The aim of this study is to characterize the different precipitates that dominate the microstructure of the AA2195 alloy. Since the composition evaluation of sub-nanometer scale platelet precipitates is extremely difficult, atom probe tomography has been employed to characterize the local chemical composition of the precipitates and to determine the location of the various alloying atoms.

## 2. Experimental:

The chemical composition of AA2195 is shown in table 1. The table shows the nominal composition for the alloy and the measured composition by inductively coupled plasma optical emission spectrometry (Varian ICP-OES 720-ES). Samples in the form of ingots were obtained in a T8 temper state. Samples on the T8 condition were solution treated for 1h to 783 K (510 ˚C), quenched, stretched to 3 % and subsequently aged to (423 K) 150 ˚C for 30 h.

Vickers micro-hardness measurements were performed using a load of 0.2 kg applied on the sheet sample. The Vickers hardness number (VHN), which was obtained by averaging ten measurements, was found to be (160±7) HV/1.96 and thus confirms the success of the heat treatment.

Small rods (0.3 mm x 0.3 mm x 10 mm) were spark machined from the bulk and then electro-polished to a sharp needle-shaped specimen for atom probe tomography (APT) analyses and scanning transmission electron microscopy (STEM). Electro-polishing was carried out using a solution of 30 vol. % nitric acid in methanol at temperatures between 253 to 298 K (-20 to 25 ˚C) in the range of 5 – 7 V .The specimens were prepared for transmission electron microscopy (TEM) by standard twin-jet electro-polishing using a solution of 30 vol. % nitric acid in methanol at 243 K (-30 ˚C).

STEM images were recorded using a Titan CT with a dedicated APT holder (E.A. Fischione 2025 Tomoholder). TEM analysis was carried out using a Jeol2100F, operated at 200 kV. The APT experiments were performed on both the Cameca laser assisted wide angle tomographic atom probe (LAWATAP) in the voltage mode at temperatures between 25 to 30 K (-248 to -243 ˚C) with a pulse fraction between 20% and 22.5% at a vacuum level of $10^{-8}$ Pa, and the Cameca local electrode atom probe (LEAP- 4000HR) in the voltage mode at 22 K (-251 °C) and 18% of pulse fraction at $10^{-8}$ Pa.

The obtained data was visualized using the TAP3D and IVAS software programs provided by Cameca. A proximity histogram (proxigram) was used to obtain the chemical composition within the precipitates [14]. This proxigram computes the composition as a function of the distance to the isoconcentration surface. All proxigrams were computed with a 0.1 nm bin size. To increase the statistical accuracy of the composition, bins containing fewer than 50 atoms were eliminated.

An isoconcentration surface is the surface that envelops a volume with concentrations higher than the selected concentration of one or more elements. This surface was obtained by sampling the atom probe tomography reconstruction with $0.8 \text{ X } 0.8 \text{ X } 0.8 \text{ nm}^3$ voxels after applying a delocalization procedure developed by Hellman et al. [15] with smoothing parameters of 1.5 nm for the x and y coordinates and 0.75 nm for the z coordinate.

**3. Results and discussion:**

Several different precipitates were identified in AA2195 and are discussed below:

3.1 Electron microscopy:

Conventional TEM bright field images revealed a complex microstructure encompassing different precipitates with platelet and spherical morphologies as shown in fig. 1. Precipitates were characterized using an exact [110] zone to detect $T_1$ platelets, $\theta'$ platelets and spherical precipitates with $L1_2$–structure. A schematic of the indexed selected area diffraction pattern (SADP) in the [101] direction is integrated with the bright field- TEM image (fig. 1(a)). This indicates the presence of two variants of $T_1$ precipitate as streaks along the <111> direction in the pattern. The other two variants of this phase produce spots adjacent to <200> positions. The streaks along the <200> direction reveal the presence of one variant of $\theta'$. The super-lattice reflections in the diffraction pattern may originate from an $L1_2$-ordered structure representing $\delta'(Al_3Li)$ and/or $\beta'(Al_3Zr)$ precipitates. The $L1_2$ unit cell is based on the face centered cubic (fcc) unit cell of aluminum. However, the corner atoms are replaced by either Li or Zr, respectively. This means that all <100> directions are superstructure directions. The spherical precipitates are most probably of the $\beta'$-type, since the Li composition in the investigated alloy is less than 5at.%. It is worth to note that from the diffraction pattern alone, no decision can be achieved whether the precipitates are $\delta'$ or $\beta'$ because of the similarity of the lattice constant. Fig. 1(b) presents a simulated diffraction pattern matching all spots and streaks in the [101] selected area diffraction pattern.

High angle annular bright field scanning transmission electron microscopy (HAABF) STEM - contrast also revealed the existence of both spherical and platelet precipitates (fig. 2). As illustrated in fig. 2, the heterogeneous formation of platelet precipitate can be clearly observed at grain boundaries by imaging the sample from different angles (± 75˚ in steps of 1˚). A low angle grain boundary in fig. 2 was identified while rotating the specimen and monitoring the diffraction pattern at various locations within the sample. Moreover, the distribution of platelet precipitates and spherical precipitates within the matrix could be observed. The heterogonous nucleation site of the $T_1$-phase on the grain boundary agrees well with the previous observation reported by Cassade et al. [6], who proposed that the precipitation of the $T_1$ phase at sub-grain boundaries results from the easy migration of Cu and Li atoms and the sufficient concentration of mobile vacancies at these locations. Moreover, Lee et al. [16] observed that an increase in aging time caused a rapid precipitation of the $T_1$ phase both within the matrix and at sub-grain boundaries, consistent with the (HAABF) STEM image presented in fig. 2.

3.2 Atom probe tomography analyses:

The presence of several alloying elements such as Li, Cu, Mg, Zr and Ag in AA2195 leads to complex interactions between the different types of precipitates, which has led to many studies on the evolution of such precipitates in alloys. The use of atom probe tomography allows for the determination of the chemical composition within small-thickness platelets (less than 2 nm) embedded in an Al matrix as discussed in this section.

3.2.1 LAWATAP analyses at 30 K (-243 ˚C) and 20% of pulse fraction:

Fig. 3 shows the reconstructed volume of the sample analyzed by a LAWATAP at 30 K (-243 ˚C) and 20% of pulse fraction (ratio of pulse voltage, $V_{pulse}$ to DC standing voltage, $V_{DC}$).

This fig. demonstrates that platelet precipitates are distributed within the matrix. Based on the knowledge of the angles between different crystallographic directions, it can be confirmed that the two plates in the upper part of the reconstructed volume are $T_1$ precipitates. The direction of the analysis has been identified as <200>. This identification is based on measuring the distance between the atomic planes in the atom probe tomography data set that are perpendicular to this zone axis, that is, the <200> direction. As shown in fig. 3(a), the angle between the normal of the $T_1$ platelet plane, the <111> direction, and the <200> zone axis was found to be (55˚±1), which is within the margin of error for the expected crystallographic angle between these two directions (the calculated value of this angle from the unit cell is 54.76˚). Due to their orientations perpendicular to the <200> zone axis, the remaining platelets in the reconstructed volume can be identified as θ′ precipitates.

A combined corresponding proxigram profile for the $T_1$ platelets is shown in fig. 3(b). The $T_1$ precipitates are delineated by 6 at.% Li isoconcentartion surface. The chemical composition for the $T_1$ platelets was found to be (11±1) at.% Cu, (13±1) at.% Li, (3±1) at.% Mg and (1.5±0.4) at.% Ag, indicating a clear deviation from the ($Al_2CuLi$) stoichiometry. A combined proxigram profile from 4 at.% Cu isoconcentration surfaces of $\theta'$ platelets is shown in fig. 3(c). According to this profile, the chemical composition for the $\theta'$ precipitates was found to be (27±3) at.% Cu, (3.8±0.2) at.% Li, and (0.7±0.2) at.% Mg. This quantitative analysis reveals a clear evidence for a significant concentration level of Li atoms inside the $\theta'$ platelets, without any indication for the presence of Ag or Zr. The chemical compositions for these platelets are very close to the ($Al_2Cu$) stoichiometry.

To obtain a high spatial resolution perpendicular to the platelets, it is desirable to align this direction with the tip axis of the prepared samples. Since the material possesses a strong rolling texture (sheet normal to {110} planes, rolling direction is ⟨112⟩), it is possible to prepare samples with a defined orientation, e.g. samples were prepared with their axis parallel to the <111> direction. Fig. 4 shows APT results for such preferentially oriented samples. The (111) Al atomic planes can be identified. Thus three out of four orientations of one $T_1$ precipitate on the {111} matrix planes can be observed (fig. 4(a)). The magnified view of one $T_1$ platelet with resolved {111} planes is shown in fig. 4(b). Fig. 4(c) is a proxigram profile from 4 at.% Cu isoconcentartion surface of the $T_1$ platelet. Even though that the measured chemical composition for this platelet were improved by preforming an APT analysis in this direction, the deviation from the stoichiometric value can clearly be seen in the chemical composition of (14±2) at.% Cu, (14±3) at.% Li, (4±1) at.% Mg and (1.4±0.4) at.% Ag. This quantitative analysis demonstrates the enrichment of Mg atoms inside the platelet without any indication for enrichment at the interface. Furthermore, there is no statistical significant enrichment of Ag or Zr. This corresponds with previous observation reported by Gault et al. [13].

The deviation of the experimentally measured chemical composition of the $T_1$ platelet to the ($Al_2CuLi$) stoichiometry has been observed by Murayama and Hono [12], as well. They interpreted this deviation as a result of the unstable ionization behavior of atoms near or within the $T_1$ precipitates resulting from the very low evaporation field of Li. Moreover, Gault et al. [13] interpret this deviation as a result of a convolution of the matrix concentration with that of the precipitates due to the presence of trajectory aberrations in the APT analyses. This artifact leads to an overestimation of the Al content inside the platelet. However, significant notes could be observed on the correction method suggested by Gault et al. [13]. Most notably the assumption of having a correct aluminum contents in all precipitates and calculating their chemical composition as a function of apparent aluminum concentration enhances the uncertainties of their measured chemical composition.

As far as the trajectory aberrations are concerned, the thickness of the platelets will be overestimated in the 3Dreconstructed volume (fig. 3(a) and 4(a)). However, an accurate platelet thickness can be determined by counting the atomic planes of the Al-rich matrix in the <111> direction. By multiplying the number of Al atomic planes by the interplanar spacing of $a/\sqrt{(h^2+k^2+l^2)}$ (where $a$ is the lattice parameter for fcc Al, $a$=0.405 nm and $h=k=l=1$) the $T_1$ platelet thickness was found to be (1.4 ±0.2) nm. As the presence of different alloying elements with low chemical concentrations [17] does not affect largely the value of the lattice parameter of the Al-rich matrix (table 1), it can be justified to set the value of the lattice parameter for the Al-rich matrix to be equal to that of pure Al. Furthermore, local magnification effects, that cause variations in atomic density between matrix and precipitates [18], can lead to an underestimation of the Li-and Cu-content of the thin $T_1$ platelets. These APT artifacts might affect the estimation of the chemical composition for the $T_1$ platelets precipitates, and hence great care is required for the analyses of this phase. Menand et al. have reported optimized experimental conditions for the analysis of Al-Li alloys which allow minimizing such APT artifacts [9]. These conditions consist of a low tip temperature of 20 to 30 K (-253 to-243 °C) and a pulse fraction of greater than 15%. On this basis, atom probe tomography experiments were slightly changed.

3.2.2 LAWATAP analyses at 25 K (-248 °C) and 22.5% of pulse fraction:

The series of images in fig. 5 reveals the effect of changing the experimental parameters on the detected chemical composition within the precipitates. Fig. 5(a) shows various $T_1$ and $\theta'$ platelets. The proxigram profiles were computed by following the same procedure as mentioned above. Proxigram profiles corresponding to $\theta'$ and $T_1$ platelets where $\theta'$ precipitates are delineated by 10 at.% Cu isoconcentartion surfaces and $T_1$ precipitates are delineated by 10 at.% Li isoconcentartion surfaces are presented in fig. 5(b) and 5(c), respectively. As illustrated in fig. 5(b), the chemical composition for $\theta'$ was found to be (30±2) at.% Cu, (3±1) at.% Li, and (1 ±0.7) at.% Mg. It is worth noting that analyses of the $\theta'$ precipitates under these experimental conditions resulted in $\theta'$ stoichiometries consistent with the expected ($Al_2Cu$) equilibrium composition.

A combined proxigram profile for several $T_1$ platelets is shown in fig. 5(c). The chemical compositions for the $T_1$ platelets was found to be (14±1) at.% Cu, (14±2) at.% Li, (3.8±1) at.% Mg and (0.8 ±0.4) at.% Ag. This confirms both the deviation from ($Al_2CuLi$) stoichiometry and a potential enrichment of the $T_1$ phase with Mg.

3.2.3 LEAP analyses at 22 K (-251 ˚C) and 18% of pulse fraction:

The use of the wider field of view of the detector in the LEAP permits the collection of larger datasets. Furthermore, using a different machine allows to determine any device specific artifacts.

A top view of the reconstructed volume of a sample analyzed by a LEAP in voltage mode at 22 K (-251 ˚C) and 18% pulse fraction is shown in fig. 6(a). The intersections of three $T_1$ platelets with the (111) pole are visible. According to the proxigram (fig. 6(b)) from 8at.% Li isoconcentration surface of these platelets, this analysis yielded the close values for the chemical compositions of the $T_1$ platelets, with (14.3±2) at.% Cu, (15.5±3) at.% Li, (2.5±1) at.% Mg and (0.8±0.5) at.% Ag, as LAWATAP analyses.

The combination of results from different atom probes at different experimental conditions, even including laser based results from Gault [13], suggest that the actual chemical composition for the thin $T_1$ platelets could differ from the ($Al_2CuLi$) of the stable bulk phase, since changing the parameters should induce a change in the significance of the artifacts. Furthermore, it seems unlikely that the artifacts would only influence the Cu detection for the $T_1$ platelets but not the $\theta'$ platelets. Even if one assumes that Li changes the evaporation behavior of Cu, it is difficult to understand why it should not also affect Al and thus negate this effect. Such an effect should also result in an extremely decreased atomic density, since about half of all otherwise detected atoms would be missing. But this is not observed at the platelet positions according to the density profiles of the cylinders perpendicular to these thin platelets. In fact, the average difference between the detected atoms in the matrix and the precipitates dose not excess 32 atoms.

In conclusion, it seems reasonable to propose that the concentration in the $T_1$ phase is indeed non stoichiometric after treatment of the samples in the T8 condition. This finding is also supported by High angle annular dark field scanning transmission electron microscopy (HAADF) STEM and small angle X-ray scattering and results from Donnadieu et al. [19]. In this study the author investigated a similar alloy, conducting a similar thermo-mechanical treatment. The observed microstructure also consisted a homogenous distribution of thin platelets precipitates with 1 nm thickness and parallel to (111) Al planes. They noted that the composition of the very thin $T_1$ precipitates may deviate from the ($Al_2CuLi$) stoichiometry of the stable bulk phase and proposed that this may be due either to the presence of higher Al content than would be expected for the $T_1$ ($Al_2CuLi$), or due to the nature of the interface planes. Their results reported on Al content higher than expected for $T_1$ ($Al_2CuLi$) phase and the occurrence of matrix/$T_1$ interface on Al-Cu mixed layer. The authors also, suggested that the actual atomic structure of the thin $T_1$ platelet could present a stoichiometry different from that for the bulk $T_1$ phase.

More recently, an APT study by V. Araullo-Peters et al. [20] reported on the presence of the $T_1$ precursor at the early stage of ageing. However, the observed $T_1$ platelets in this study are between 5 and 10 nm in thickness. Their reported results showed a different chemical composition for these $T_1$ platelets from that for the bulk $T_1$ phase.

In particular, our results demonstrated the presence of lower Li and Cu contents with 1:1 ratio in the thin $T_1$ platelets. These results confirm the fact of the higher Al content than would be expected for this phase. As far as the matrix/$T_1$ interface considered, APT is the more appropriate tool to study such effect. Fig. 7 displays a higher magnification image for the $T_1$ platelet. It can be seen that the density of Cu atoms at the matrix/$T_1$ interface is higher than that for Li atoms. Moreover, the correlation between Mg and Li atoms is clearly visible. This observation confirm the suggested fact by Donnadieu et al. [19] of the occurrence of Al-Cu interface instead of Al-Li interface which suggests that the first compact Li layer is involved in the nucleation of the $T_1$ phase. The deviation of the $T_1$ phase from thermodynamic equilibrium composition for the stable bulk phase suggests that the T8 condition still has appreciable supersaturation.

## 4. Conclusion:

This paper reports a characterization of the commercial aluminum-lithium-copper alloy AA2195 in the T8 condition using electron microscopy and atom probe tomography. The results confirm that the microstructure of the alloy in this temper state consists of platelet-shaped $T_1(Al_2CuLi)$, $\theta'(Al_2Cu)$ and spherical $\beta'(Al_3Zr)$ precipitates.

Owing to the ability of atom probe tomography to resolve the chemical position on a sub-nanometer scale, compositional analyses of thin platelet precipitates (less than 2 nm) have been performed. The measured $\theta'$ stoichiometry is consistent with the expected $(Al_2Cu)$ equilibrium composition with a significant partitioning of Li atoms within this precipitate. For the $T_1$ platelets, a deviation from the stoichiometric $(Al_2CuLi)$ of the bulk phase was observed. This phase is enriched with Mg atoms without any indication of the presence of Ag atoms at the phase/matrix interface. The off-stoichiometry of the thin $T_1$ platelet cannot be simply attributed to artifacts in the APT measurements and might be considered as a true deviation from the thermodynamic equilibrium composition of the thin $T_1$ platelets for the stable bulk phase.

**Acknowledgments:**

The Authors thank Dr. Dalaver H. Anjum (Imaging and characterization core lab – KAUST) for his support with electron tomography. F. Haider acknowledges financial support provided to him during his visits at King Abdullah University of Science and Technology (KAUST). T. Al-Kassab gratefully acknowledges financial support provided to him through King Abdullah University of Science and Technology (KAUST) base-line funding program.

**References:**

[1] J. R. Pickens, F. H. Heubaum, T. J. Langan and L. S. Kramer: in Aluminum-Lithium Alloys (proc. of the fifth int. Al-Li Conf.) (edited by T. H. Sanders and E. A. Starke), p. 1397, MCE publications Ltd., Birmingham, U.K, 1989.

[2] E. A. Jr Starke and B. N Bhatm: Technical Summary in: Aluminum-Lithium Alloys for Aerospace Applications Workshop, (eds B. N. Bhat, T. T. Bales and E. J. Vesely, Jr ) , pp. 3-5, NASA Conference Publication 3287, Marshall Space Flight Center, Alabama, 1994.

[3] R. Crooks, Z. Wang, V. I. Levit and R.N. Shenoy: Mater. Sci. Eng., A, 1998, Vol. 257(1), pp. 145-152.

[4] P.S. Chen and B.N. Bhat: Report No. 211548, NASA/TM, Alabama, 2002.

[5] G. Itoh, Q. Cui and M. Kanno: Mater. Sci. Eng., A, 1996, Vol. 211(1–2), pp. 128-137.

[6] W.A. Cassada, G.J. Shiflet, and E.A. Starke: Metall. Trans. A, 1991, Vol. 22(2), pp. 299-306.

[7] C. Laird and H.I. Aaronson : Trans Metall Soc AIME, 1968, Vol. 242, pp. 1393.

[8] E. A. Jr Starke and J. T. Staley: Prog. Aerospace Sci, 1995, Vol. 32, pp. 131-172.

[9] A. Menand, T. Al-Kassab, S. Chambreland and J.M. Sarrau: J. de Phys, 1988, Vol. 49. pp. 353.

[10] L. H. A. Terrones and S. N. Monteir: Mater. Charact, 2006, Vol. 58(2), p. 156-161.

[11] G. Schmitz, K. Hono and P. Haasen: Acta. Metal. Mater, 1993, Vol. 42(1), p. 201-211.

[12] M. Murayama and K. Hono: Scripta Mater., 2001, Vol. 44, pp. 701-706.

[13] B. Gault, F. de Geuser, L. Bourgeois, B.M. Gabble, S.P. Ringer and B.C. Muddle: Ultramicroscopy, 2011, Vol. 111(6), pp. 683-689.

[14] O.C. Hellman, J.A. Vandenbroucke, J. Rusing, D. Isheim and D.N. Seidman: Microsc. Microanal, 2000, Vol. 6(5), pp. 437-444.

[15] O.C. Hellman, J.B. Du Rivage and D.N. Seidman: Ultramicroscopy, 2003, Vol. 95, pp. 199-205.

[16] K. H. Lee, Y. J. Lee, and K. Hiraga: J. Mater. Res, 1998. Vol. 14, No. 2.

[17] S. H. Kellington, D. Loveridge and J. M. Titman: BRIT. J. APPL. PHYS., 1969, Vol. 2.

[18] F. Vurpillot, A. Bostel, A. Menand and D. Blavette: Eur Phys J-Appl Phys, 1999, Vol. 6, pp 217-221.

[19] P. Donnadieu, Y. Shao, F. De Geuser, G. A. Botton, S. Lazar, M. Cheynet, M. de Boissieu and A. Deschamps : Acta Mater., 2011, Vol. 59(2), pp. 462-472.

[20] V. Araullo-Peters, B. Gault, F. De Geuser, A. Deschamps and J. M. Cairney: Acta Mater., 2014, Vol. 66, pp. 199-208.

Table 1. Chemical composition of alloying elements in AA2195.

| Alloying element | Cu | Li | Mg | Ag | Zr | Al |
|---|---|---|---|---|---|---|
| Nominal (wt. %) | 4 | 1 | 0.36 | 0.28 | 0.14 | 94.22 |
| Nominal (at. %) | 1.69 | 3.87 | 0.39 | 0.06 | 0.04 | 93.95 |
| ICP (at. %) | 1.74±0.19 | 3.67±0.05 | 0.36±0.01 | 0.08±0.02 | 0.04±0.0004 | 94.11±0.27 |

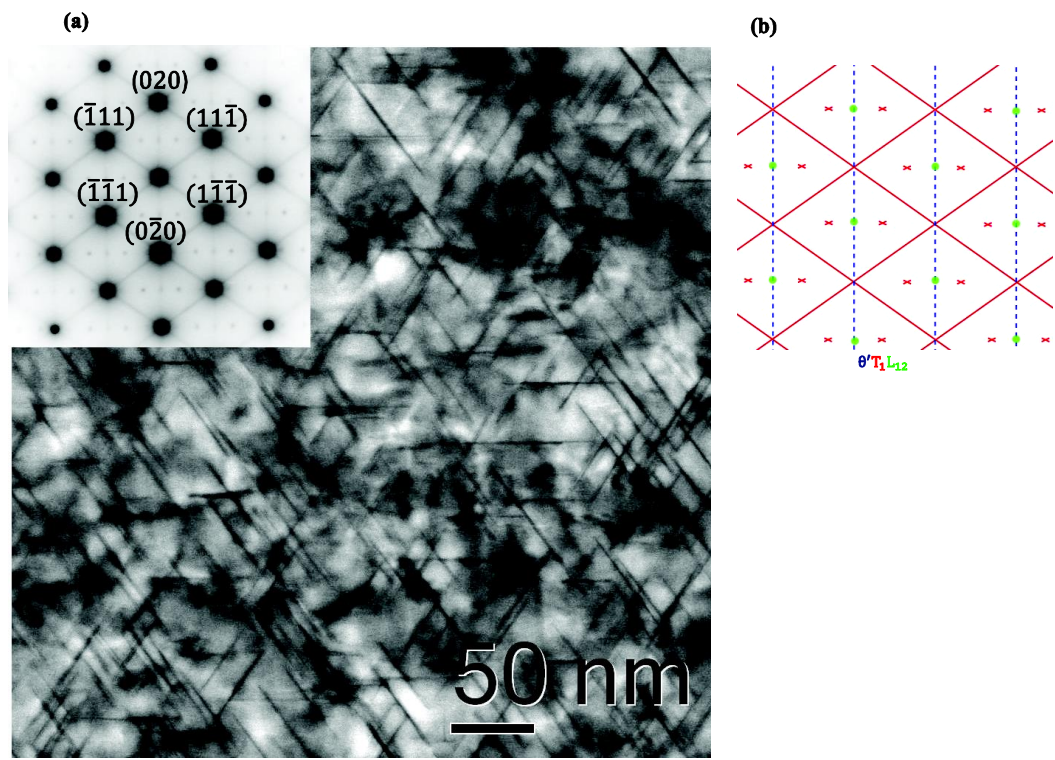

Fig. 1. Evolution of the microstructure as observed by TEM. (a) Typical bright field image showing a complex microstructure with the $T_1$, $\theta'$ and $\beta'$ precipitates and the corresponding [101] selected area diffraction pattern (SADP). (b) Schematic of a simulated diffraction pattern that assigns the reflections and streaks in the SADP to the corresponding precipitates.

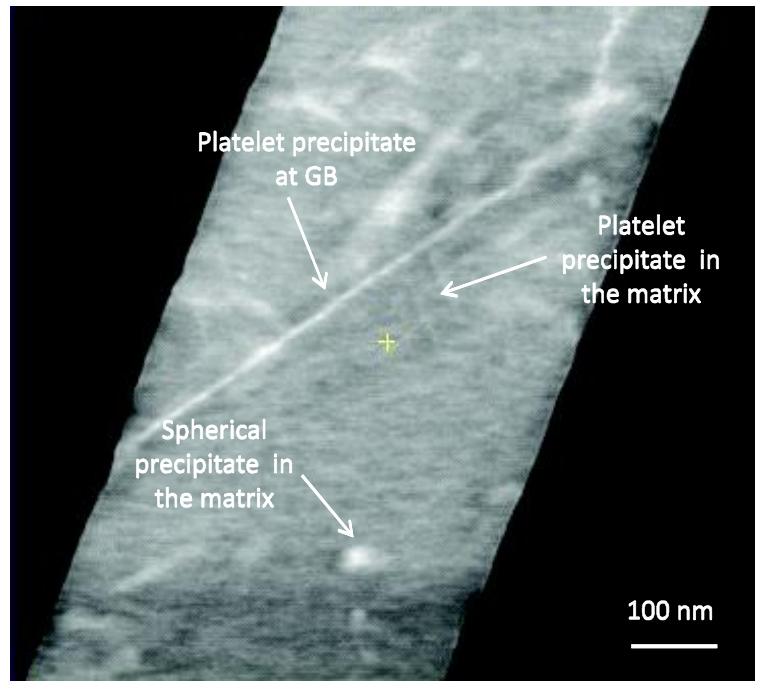

Fig. 2. High angle annular bright field scanning transmission electron microscopy image showing the heterogeneous formation of a bright imaging platelet precipitate at the grain boundary and the distribution of platelets and spherical precipitates within the matrix.

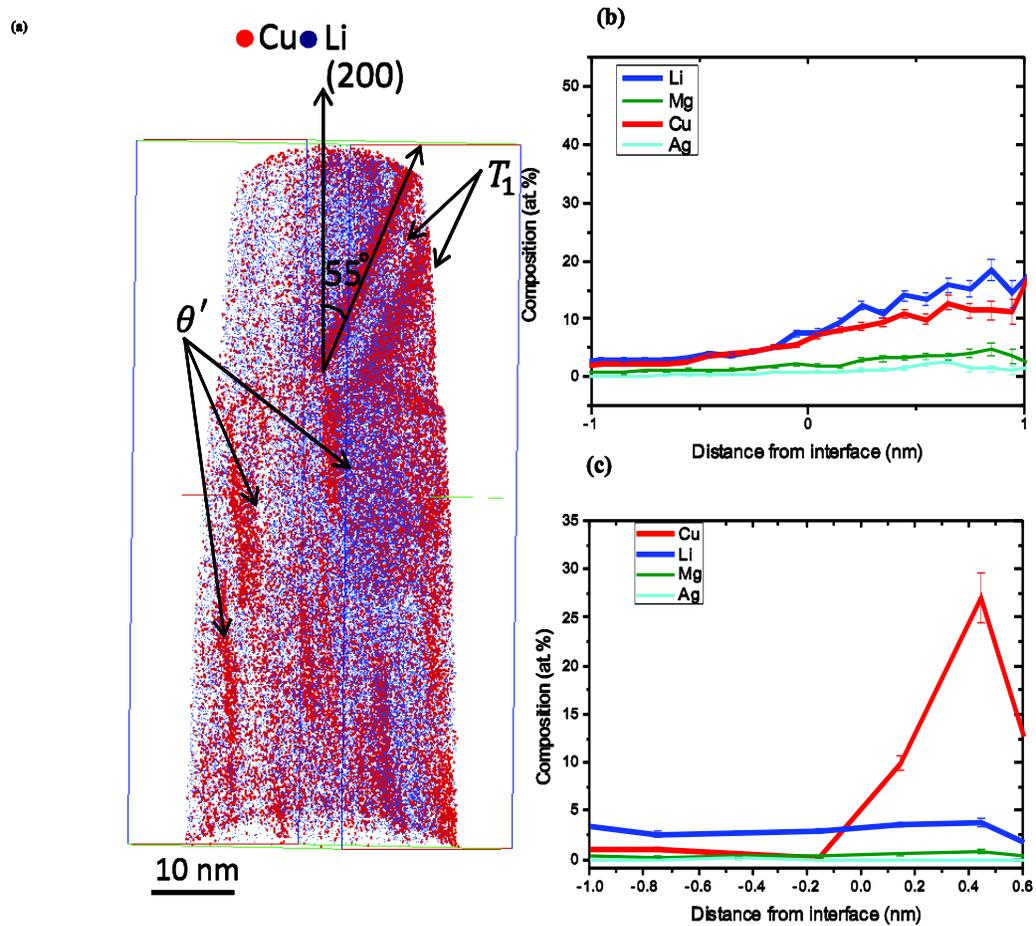

Fig. 3. LAWATAP analysis of the microstructure at 30 K (-243 °C) and 20% of pulse fraction. (a) Reconstructed volume showing the distribution of the $T_1$ and $\theta'$ phases. (b) Corresponding proxigram composition profile for the $T_1$ platelet indicating a chemical composition of (11±1) at.% Cu, (13±1) at.% Li, (3±1) at.% Mg and (1.5±0.4) at.% Ag (c) Corresponding combined proxigram composition profile for the $\theta'$ platelet with a chemical composition of (27±3) at.% Cu, (3.8±0.2) at.% Li, and (0.7 ±0.2) at.% Mg.

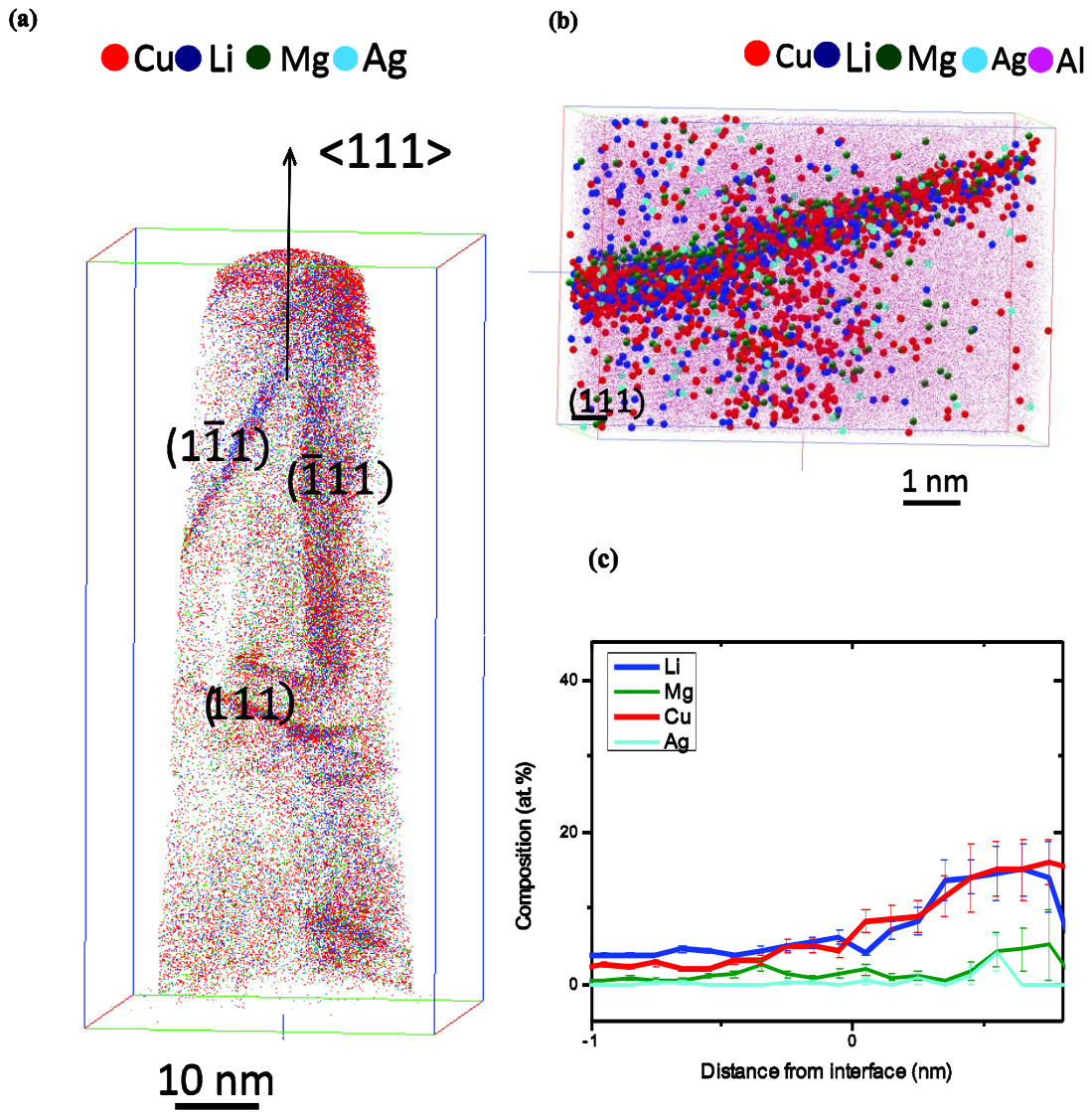

Fig. 4. LAWATAP analysis for the specimen with rolling texture with the same experimental parameters as in fig. 3. (a) Reconstructed volume showing three out of four orientations of the $T_1$ precipitate on the {111} matrix planes. (b) Magnified view of the $T_1$ platelet with resolved {111} planes reveals the distribution of Mg within the $T_1$ platelet. (c) Corresponding proxigram composition profile for the $T_1$ platelet on the {111} plane gives the chemical composition as (14±2) at.% Cu, (14±3) at.% Li, (4±1) at.% Mg and (1.4±0.4) at.% Ag.

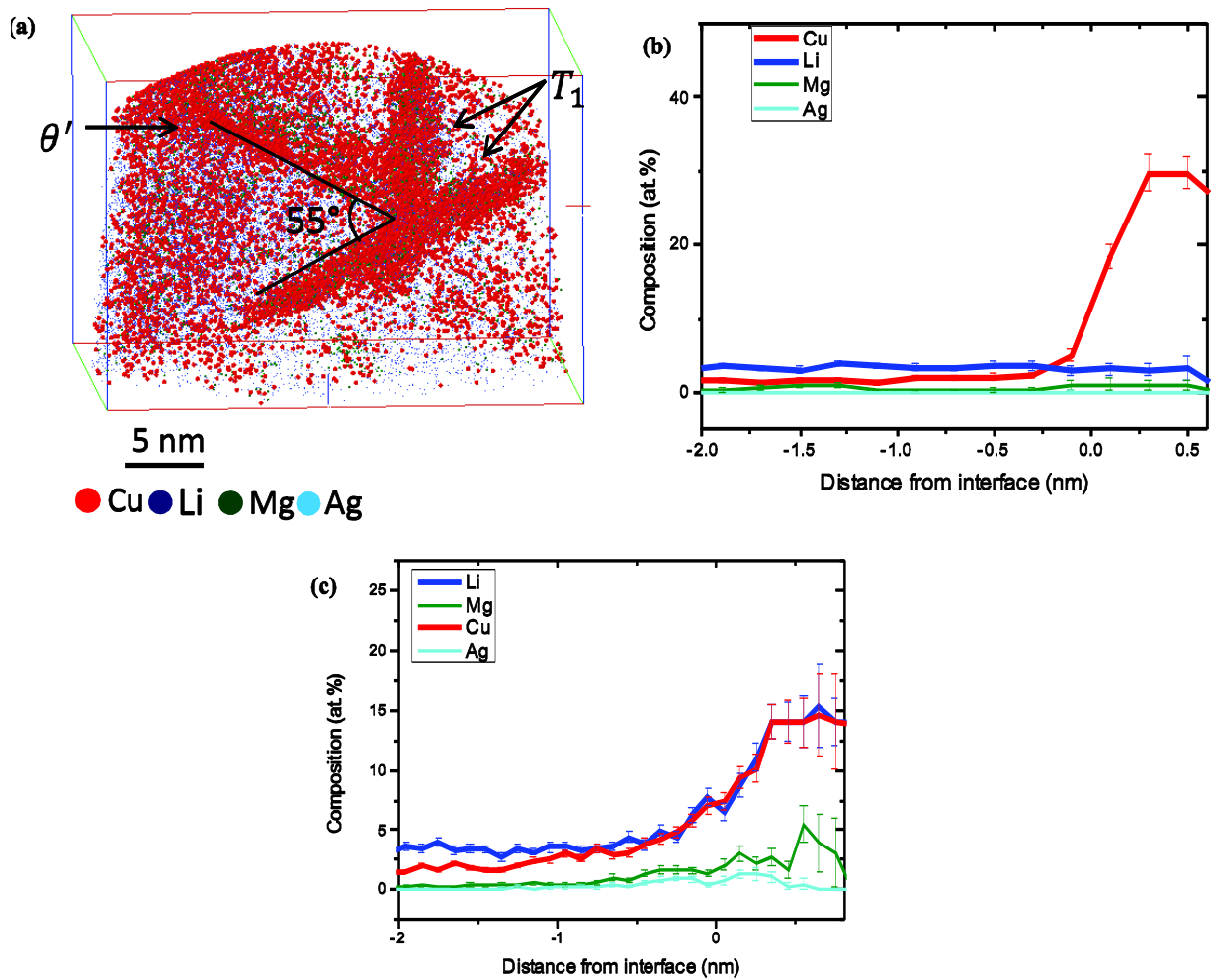

Fig. 5. LAWATAP analysis with new experimental parameters (25 K (-248 °C) and 22.5% of pulse fraction). (a) Reconstructed volume, obtained with laser assisted wide angle tomographic atom probe, shows $T_1$ and $\theta'$ platelets. (b) Corresponding proxigram composition profile for $\theta'$ gives an optimistic chemical composition as (30±2) at.% Cu, (3±1) at.% Li, and (1±0.7) at.% Mg. (c) Corresponding combined proxigram composition profile for $T_1$ with the chemical composition of (14±1) at.% Cu, (14±2) at.% Li, (3.8±1) at.% Mg and (0.8 ±0.4) at.% Ag.

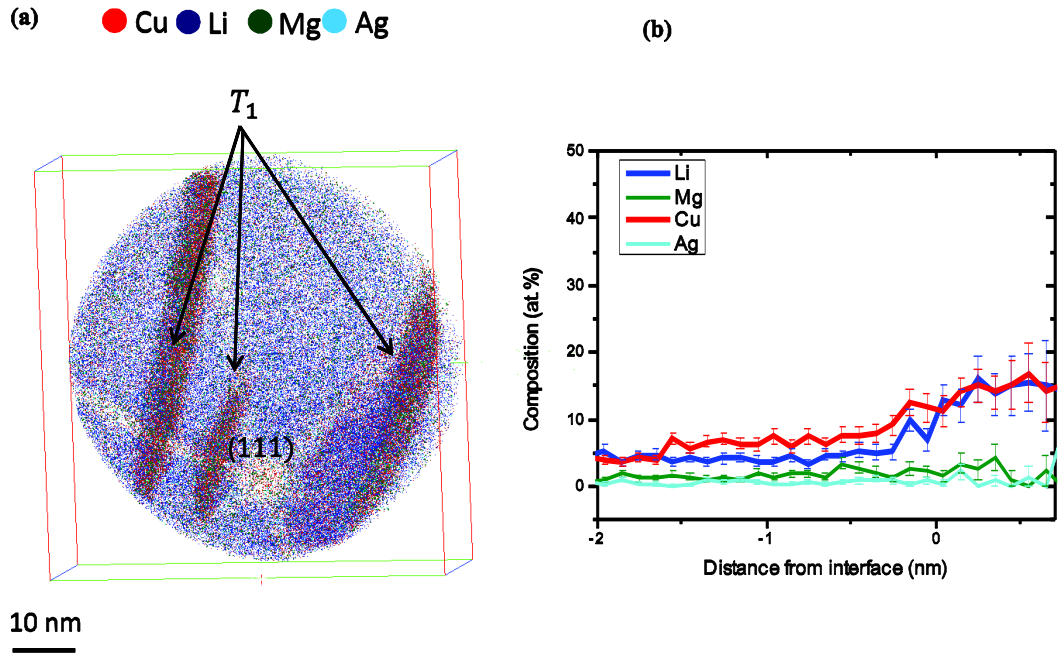

Fig. 6. LEAP analysis at 22 K (-251 °C) and 18% of pulse fraction. (a) Top view of the reconstructed volume of a sample shows the intersections of the $T_1$ platelets with the (111) pole. (b) Corresponding combined proxigram composition profile for $T_1$ with the chemical composition of (14.3±2) at.% Cu, (15.5±3) at.% Li, (2.5±1) at.% Mg and (0.8±0.5) at.% Ag.

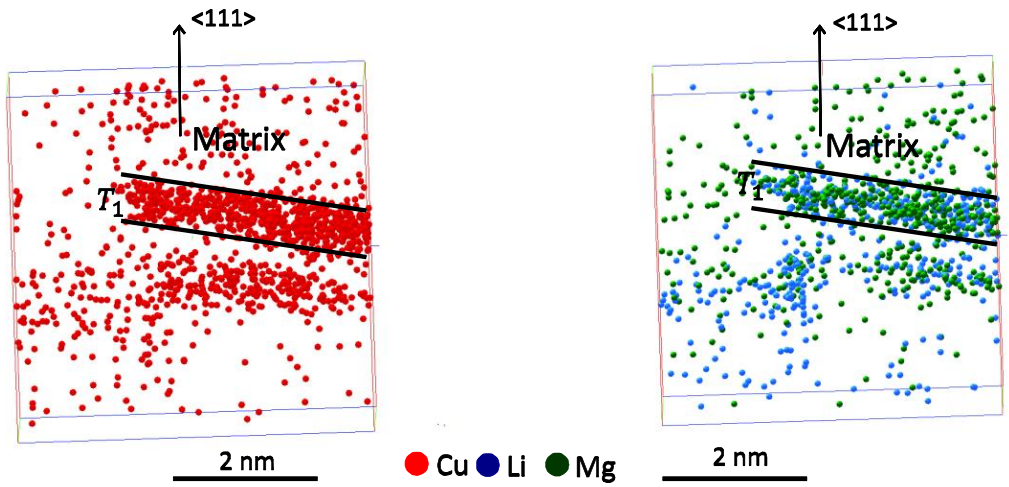

Fig. 7. The magnified view of the $T_1$ platelet showing the density of the Cu, Li and Mg atoms at matrix/ $T_1$ interface.